# Using the No-Search Easy-Hard Technique for Downward Collapse[*]

Edith Hemaspaandra[†]
Department of Computer Science
Rochester Institute of Technology
Rochester, NY 14623, USA

Lane A. Hemaspaandra[‡]
Department of Computer Science
University of Rochester
Rochester, NY 14627, USA

Harald Hempel[§]
Institut für Informatik
Friedrich-Schiller-Universität Jena
07743 Jena, Germany

June 13, 2001

**Abstract:**

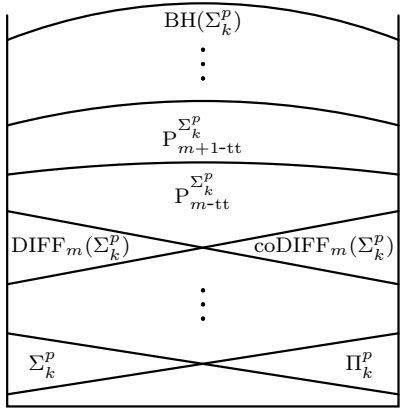

---

[*]Supported in part by grants NSF-CCR-9322513, NSF-INT-9513368/DAAD-315-PRO-fo-ab, and NSF-INT-9815095/DAAD-315-PPP-gü-ab, and a NATO Postdoctoral Science Fellowship from the DAAD's "Gemeinsames Hochschulsonderprogramm III von Bund und Ländern" program. This report combines and extends each of technical reports corr-cs.CC/9808002 and corr-cs.CC/9910008.

[†]Email: `eh@cs.rit.edu`. Work done in part while visiting Friedrich-Schiller-Universität and Julius-Maximilians-Universität.

[‡]Email: `lane@cs.rochester.edu`. Work done in part while visiting Friedrich-Schiller-Universität and Julius-Maximilians-Universität.

[§]Email: `hempel@informatik.uni-jena.de`. Work done in part while visiting Le Moyne College, and while visiting the University of Rochester under a NATO Postdoctoral Science Fellowship from the DAAD.

The top part of the preceding figure shows some classes from the (truth-table) bounded-query and boolean hierarchies. It is well-known that if either of these hierarchies collapses at a given level, then all higher levels of that hierarchy collapse to that same level. This is a standard "upward translation of equality" that has been known for over a decade. The issue of whether these hierarchies can translate equality *downwards* has proven vastly more challenging. In particular, with regard to the figure above, consider the following claim:

$$P^{\Sigma_k^p}_{m\text{-tt}} = P^{\Sigma_k^p}_{m+1\text{-tt}} \Rightarrow \text{DIFF}_m(\Sigma_k^p) = \text{coDIFF}_m(\Sigma_k^p) = \text{BH}(\Sigma_k^p). \quad (*)$$

This claim, if true, says that equality translates downwards between levels of the bounded-query hierarchy and the boolean hierarchy levels that (before the fact) are immediately below them.

Until recently, it was not known whether (*) *ever* held, except for the degenerate cases $m = 0$ and $k = 0$. Then Hemaspaandra, Hemaspaandra, and Hempel [13] proved that (*) holds for all $m$, for $k > 2$. Buhrman and Fortnow [4] then showed that, when $k = 2$, (*) holds for the case $m = 1$. In this paper, we prove that for the case $k = 2$, (*) holds for all values of $m$. Since there is an oracle relative to which "for $k = 1$, (*) holds for all $m$" fails [4], our achievement of the $k = 2$ case cannot to be strengthened to $k = 1$ by any relativizable proof technique. The new downward translation we obtain also tightens the collapse in the polynomial hierarchy implied by a collapse in the bounded-query hierarchy of the second level of the polynomial hierarchy.

# 1  Introduction

Does the collapse of low-complexity classes imply the collapse of higher-complexity classes? Does the collapse of high-complexity classes imply the collapse of lower-complexity classes? These questions—known respectively as upward and downward translation of equality—have long been central topics in computational complexity theory. For example, in the seminal paper on the polynomial hierarchy, Meyer and Stockmeyer [18] proved that the polynomial hierarchy displays upward translation of equality (e.g., $P = NP \Rightarrow P = PH$).

The issue of whether the polynomial hierarchy—its levels and/or bounded access to its levels—ever displays *downward* translation of equality has proven more difficult. The first such result regarding bounded access was recently obtained by Hemaspaandra, Hemaspaandra, and Hempel [13], who proved that if for some high level of the polynomial hierarchy one query equals two queries, then the hierarchy collapses down not just to one query to that level, but rather to that level itself. That is, they proved the following result (note: the levels of the polynomial hierarchy [18, 24] are denoted in the standard way, namely, $\Sigma_0^p = P$, $\Sigma_1^p = NP$, $\Sigma_k^p = NP^{\Sigma_{k-1}^p}$ for each $k > 1$, and $\Pi_k^p = \{L \mid \overline{L} \in \Sigma_k^p\}$ for each $k \geq 0$).

**Theorem 1.1** *([13])   For each $k > 2$: If $P^{\Sigma_k^p[1]} = P^{\Sigma_k^p[2]}$, then $\Sigma_k^p = \Pi_k^p = PH$.*

This theorem has two clear directions in which one might hope to strengthen it. First, one might ask not just about one-versus-two queries but rather about $m$-versus-$m+1$ queries. Second, one might ask if the $k > 2$ can be improved to $k > 1$. Both of these have been achieved. The first strengthening was achieved in a more technical section of the same paper by Hemaspaandra, Hemaspaandra, and Hempel [13]. They showed that Theorem 1.1 was just the $m = 1$ special case of a more general downward translation result they established, for $k > 2$, between bounded access to $\Sigma_k^p$ and the boolean hierarchy over $\Sigma_k^p$. The second type of strengthening was achieved by Buhrman and Fortnow [4], who showed that Theorem 1.1 holds even for $k = 2$, but who also showed that no relativizable technique can establish Theorem 1.1 for $k = 1$.

Neither of the results or proofs just mentioned is broad enough to achieve both strengthenings simultaneously. In this paper we derive new results strong enough to achieve this (and more). In particular, we unify and extend all the above results, and from our more general results it easily follows that (see Corollary 5.1):

For each $m > 0$ and each $k > 1$ it holds that:
$$\mathrm{P}_{m\text{-tt}}^{\Sigma_k^p} = \mathrm{P}_{m+1\text{-tt}}^{\Sigma_k^p} \;\Rightarrow\; \mathrm{DIFF}_m(\Sigma_k^p) = \mathrm{coDIFF}_m(\Sigma_k^p).$$

In particular, we obtain for the first time the cases $(k = 2 \wedge m = 2)$, $(k = 2 \wedge m = 3)$, $(k = 2 \wedge m = 4)$, and so on. As shown near the end of Section 5, the stronger downward translation we obtain yields a strengthened collapse of the polynomial hierarchy under the assumption of a collapse in the bounded-query hierarchy over $\mathrm{NP}^{\mathrm{NP}}$. (Throughout this paper, we mean the *truth-table* bounded-query hierarchy when we say bounded-query hierarchy. However, as mentioned in Section 5, our results equally well strengthen the collapse of the polynomial hierarchy under the assumption of a collapse in the *Turing* bounded-query hierarchy over $\mathrm{NP}^{\mathrm{NP}}$.)

The results that lead to the result mentioned above (i.e., Corollary 5.1) are themselves examples of downward translation of equality. Those intermediate results that are of interest in their own right are proven in Sections 3 and 4.

We conclude this section with some comments and literature pointers. As to techniques, to study *upward* translations of equality resulting from collapses of the boolean hierarchy, Kadin [15] introduced what is known as the "easy-hard technique," and that technique was employed and strengthened in a long series of papers by many authors (see the survey [12]). In particular, Hemaspaandra, Hemaspaandra, and Hempel [13] achieved Theorem 1.1 by introducing what can be called the *no-search easy-hard technique*—basically, they show that a complexity-raising search seemingly central in the easy-hard technique can in fact be eliminated, and this opens the door to downward translation of equality results for bounded query classes. The present paper builds on the no-search easy-hard technique and on a variation on that used by Buhrman and Fortnow to prove the 1-versus-2-queries case at the second level of the polynomial hierarchy. However, these approaches seem not to be strong enough to yield our result, and so we also must combine with these techniques a new approach extending beyond 1-versus-2 queries.[1] We mention that Chang has obtained some exciting applications of easy-hard-type arguments in the context of the study of approximation [7]. We also mention that there is a body of literature showing that equality of exponential-time classes translates downwards in a limited sense: Relationships are obtained connecting such to whether *sparse* sets collapse within lower time classes ([9,10], see also [6,19]; limitations of this line are presented in [1,2,14]). Other than being an interesting restricted type of downward translation of equality, that body of work has no close connection with the present paper due to that body of work being applicable only to sparse sets.

## 2 Preliminaries

To explain exactly what we do and how it extends previous results, we now state the previous results in the more general forms in which they were actually established, though in some cases with different notation or statements (see, e.g., the interesting recent paper of Wagner [26] regarding the relationship between "delta notation" and truth-table classes). Before stating the results, we very briefly remind the reader of some definitions and notations, namely the $\Delta$ levels of the polynomial hierarchy, truth-table access, symmetric difference classes, and boolean hierarchies. A detailed introduction to the boolean hierarchy, including its motivation and applications, can be found in [5,6].

**Definition 2.1**   1. *As is standard, for each $k \geq 1$, $\Delta_k^p$ denotes $\mathrm{P}^{\Sigma_{k-1}^p}$ [18]. As is standard, for*

---

[1] Regarding this new approach (and this footnote is aimed primarily at those already familiar with the techniques of the previous papers on the no-search easy-hard technique): In the previous work extending Theorem 1.1 to the boolean hierarchy (part 1 of Theorem 2.4), the "coordination" difficulties presented by the fact that boolean hierarchy sets are in effect handled via collections of machines were resolved via using certain lexicographically extreme objects as clear signposts to signal machines with (see [13, Section 3]). In the current stronger context that approach fails. Instead, we integrate into the structure of no-search easy-hard-technique proofs (especially those of [13,4]) the so-called "telescoping" normal form possessed by the boolean hierarchy over $\Sigma_k^p$ (for each $k$, see [16,5,11,27]). (The telescoping normal form guarantees that if $L \in \mathrm{DIFF}_m(\Sigma_k^p)$, then there are sets $L_1, L_2, \ldots, L_m \in \Sigma_k^p$ such that $L = L_1 - (L_2 - (L_3 - \cdots (L_{m-1} - L_m) \cdots))$ and $L_1 \supseteq L_2 \supseteq \cdots \supseteq L_{m-1} \supseteq L_m$.) This normal form has in different contexts proven useful in the study of boolean hierarchies (see, e.g., [5,6,16]), and has been used by Rohatgi in the context of a paper using the original (i.e., the with-search version of the) easy-hard technique [21].

each $m \geq 0$ and each set $A$, $\mathrm{P}^A_{m\text{-tt}}$ denotes the class of languages accepted by deterministic polynomial-time machines allowed $m$ truth-table (i.e., non-adaptive) queries to $A$ (see [17]). For each $m \geq 0$ and each complexity class $\mathcal{C}$, $\mathrm{P}^{\mathcal{C}}_{m\text{-tt}}$ is defined as $\bigcup_{A \in \mathcal{C}} \mathrm{P}^A_{m\text{-tt}}$.

2. For any classes $\mathcal{C}$ and $\mathcal{D}$, $\mathcal{C}\boldsymbol{\Delta}\mathcal{D} = \{L \,|\, (\exists C \in \mathcal{C})(\exists D \in \mathcal{D})[L = C \Delta D]\}$, where $C \Delta D = (C - D) \cup (D - C)$. We will refer to classes defined via $\boldsymbol{\Delta}$ as symmetric difference classes.

3. [5,6] Let $\mathcal{C}$ be any complexity class. The levels of the boolean hierarchy are defined as follows.

   (a) $\mathrm{DIFF}_1(\mathcal{C}) = \mathcal{C}$.

   (b) For all $m \geq 1$, $\mathrm{DIFF}_{m+1}(\mathcal{C}) = \{L \,|\, (\exists L_1 \in \mathcal{C})(\exists L_2 \in \mathrm{DIFF}_m(\mathcal{C}))[L = L_1 - L_2]\}$.

   (c) For all $m \geq 1$, $\mathrm{coDIFF}_m(\mathcal{C}) = \{L \,|\, \overline{L} \in \mathrm{DIFF}_m(\mathcal{C})\}$.

   (d) $\mathrm{BH}(\mathcal{C})$, the boolean hierarchy over $\mathcal{C}$, is $\bigcup_{m \geq 1} \mathrm{DIFF}_m(\mathcal{C})$.

The relationship between the levels of the boolean hierarchy over $\Sigma^p_k$, bounded access to $\Sigma^p_k$, and various symmetric difference classes is as follows.

**Proposition 2.2**  1. ([25])  For each $k \geq 1$ and each $m \geq 1$, $\mathrm{P}^{\Sigma^p_k}_{m\text{-tt}} \substack{\subseteq \mathrm{DIFF}_{m+1}(\Sigma^p_k) \subseteq \\ \subseteq \mathrm{coDIFF}_{m+1}(\Sigma^p_k) \subseteq} \mathrm{P}^{\Sigma^p_k}_{m+1\text{-tt}}$.

2. ([16])  For all $k \geq 1$ and all $m \geq 1$, $\mathrm{DIFF}_m(\Sigma^p_k) = \underbrace{\Sigma^p_k \boldsymbol{\Delta} \Sigma^p_k \boldsymbol{\Delta} \cdots \boldsymbol{\Delta} \Sigma^p_k}_{m \text{ times}}$.

3. ([26])  For all $k \geq 1$ and all $m \geq 1$, $\mathrm{P}^{\Sigma^p_k}_{m\text{-tt}} = \mathrm{P}\boldsymbol{\Delta}\mathrm{DIFF}_m(\Sigma^p_k) = \mathrm{P}\boldsymbol{\Delta}\underbrace{\Sigma^p_k \boldsymbol{\Delta} \Sigma^p_k \boldsymbol{\Delta} \cdots \boldsymbol{\Delta} \Sigma^p_k}_{m \text{ times}}$.

Regarding symmetric difference classes, we point out an immediate, but in the context of this paper useful, observation.

**Observation 2.3** Let $\mathcal{C}_1$, $\mathcal{C}_2$, and $\mathcal{D}$ be complexity classes. If $\mathcal{C}_1 \subseteq \mathcal{C}_2$, then $\mathcal{C}_1 \boldsymbol{\Delta} \mathcal{D} \subseteq \mathcal{C}_2 \boldsymbol{\Delta} \mathcal{D}$.

Now we can state what the earlier papers achieved (and, in doing so, those papers obtained as corollaries the results attributed to them in the Introduction).

**Theorem 2.4**  1. ([13])  Let $m > 0$, $0 \leq i < j < k$, and $i < k - 2$. If $\mathrm{P}^{\Sigma^p_i}_{1\text{-tt}}\boldsymbol{\Delta}\mathrm{DIFF}_m(\Sigma^p_k) = \mathrm{P}^{\Sigma^p_j}_{1\text{-tt}}\boldsymbol{\Delta}\mathrm{DIFF}_m(\Sigma^p_k)$, then $\mathrm{DIFF}_m(\Sigma^p_k) = \mathrm{coDIFF}_m(\Sigma^p_k)$.

2. ([4])  If $\mathrm{P}\boldsymbol{\Delta}\Sigma^p_2 = \mathrm{NP}\boldsymbol{\Delta}\Sigma^p_2$, then $\Sigma^p_2 = \Pi^p_2 = \mathrm{PH}$.

In this paper, we unify both of the above results—and achieve the strengthened corollary alluded to in the Introduction (and stated later as Corollary 5.1) regarding the relative power of $m$ and $m+1$ queries to $\Sigma_k^p$—by proving the following downward translation of equality result:

> Let $m > 0$ and $0 < i < k$. If $\Delta_i^p\boldsymbol{\Delta}\mathrm{DIFF}_m(\Sigma_k^p) = \Sigma_i^p\boldsymbol{\Delta}\mathrm{DIFF}_m(\Sigma_k^p)$, then $\mathrm{DIFF}_m(\Sigma_k^p) = \mathrm{coDIFF}_m(\Sigma_k^p)$.

## 3 A New Downward Translation of Equality

We first need a definition and a useful lemma.

**Definition 3.1** *For any sets $C$ and $D$, $C\tilde{\Delta}D = \{\langle x,y\rangle \mid x \in C \Leftrightarrow y \notin D\}$.*

**Lemma 3.2** *If $C$ is $\leq_\mathrm{m}^p$-complete for $\mathcal{C}$ and $D$ is $\leq_\mathrm{m}^p$-complete for $\mathcal{D}$, then $C\tilde{\Delta}D$ is $\leq_\mathrm{m}^p$-hard for $\mathcal{C}\boldsymbol{\Delta}\mathcal{D}$.*

**Proof:** Let $L \in \mathcal{C}\boldsymbol{\Delta}\mathcal{D}$. We need to show that $L \leq_\mathrm{m}^p C\tilde{\Delta}D$. Let $\widehat{C} \in \mathcal{C}$ and $\widehat{D} \in \mathcal{D}$ be such that $L = \widehat{C}\boldsymbol{\Delta}\widehat{D}$. Let $\widehat{C} \leq_\mathrm{m}^p C$ by $f_C$, and $\widehat{D} \leq_\mathrm{m}^p D$ by $f_D$. Then

1. $x \in L$ iff $x \in \widehat{C}\boldsymbol{\Delta}\widehat{D}$,
2. $x \in \widehat{C}\boldsymbol{\Delta}\widehat{D}$ iff $(x \in \widehat{C} \Leftrightarrow x \notin \widehat{D})$,
3. $(x \in \widehat{C} \Leftrightarrow x \notin \widehat{D})$ iff $(f_C(x) \in C \Leftrightarrow f_D(x) \notin D)$, and
4. $(f_C(x) \in C \Leftrightarrow f_D(x) \notin D)$ iff $\langle f_C(x), f_D(x)\rangle \in C\tilde{\Delta}D$.

It follows that $x \in L$ iff $\langle f_C(x), f_D(x)\rangle \in C\tilde{\Delta}D$. ∎

We now state our main result. (Note that as both $\Delta_i^p$ and $\Sigma_i^p$ contain both $\emptyset$ and $\Sigma^*$, it is clear that the classes involved in the first equality below are at least as large as the classes involved in the second equality below.)

**Theorem 3.3** *Let $m > 0$ and $0 < i < k$. If $\Delta_i^p\boldsymbol{\Delta}\mathrm{DIFF}_m(\Sigma_k^p) = \Sigma_i^p\boldsymbol{\Delta}\mathrm{DIFF}_m(\Sigma_k^p)$, then $\mathrm{DIFF}_m(\Sigma_k^p) = \mathrm{coDIFF}_m(\Sigma_k^p)$.*

This result almost follows from the forthcoming Theorem 4.1—or, to be more accurate, most of its cases are easy corollaries of Theorem 4.1. The $s = 1$ case of Theorem 4.1 states that for all $m > 0$ and all $i$ and $k$ such that $0 < i < k-1$, if $\Sigma_i^p\boldsymbol{\Delta}\mathrm{DIFF}_m(\Sigma_k^p)$ is closed under complementation, then $\text{DIFF}_m(\Sigma_k^p) = \text{coDIFF}_m(\Sigma_k^p)$. Since $\Delta_i^p \boldsymbol{\Delta} \mathcal{C}$ is closed under complementation for all $\mathcal{C}$ and all $i \geq 0$, we have that if $\Delta_i^p \boldsymbol{\Delta} \text{DIFF}_m(\Sigma_k^p) = \Sigma_i^p \boldsymbol{\Delta} \text{DIFF}_m(\Sigma_k^p)$ then $\Sigma_i^p \boldsymbol{\Delta} \text{DIFF}_m(\Sigma_k^p)$ is closed under complementation. Thus, Theorem 4.1 certainly implies Theorem 3.3 for all $m > 0$ and all $i$ and $k$ such that $0 < i < k - 1$. It remains to establish the missing cases, and Theorem 3.4 below does exactly that.

**Theorem 3.4** *Let $m > 0$ and $k > 1$. If $\Delta_{k-1}^p \boldsymbol{\Delta} \text{DIFF}_m(\Sigma_k^p) = \Sigma_{k-1}^p \boldsymbol{\Delta} \text{DIFF}_m(\Sigma_k^p)$, then $\text{DIFF}_m(\Sigma_k^p) = \text{coDIFF}_m(\Sigma_k^p)$.*

Before proving Theorem 3.4, we fix some sets that will be useful, and establish names that we will use globally for these fixed sets. (In light of the standard quantifier characterization of the polynomial hierarchy's levels [28] and the legality of padding sets to get new sets for which linear-bounded quantification suffices, it is not hard to see that there exist sets having the following properties.)

**Notation 3.5** *For each $k > 1$, fix a set $L'_{\Sigma_k^p}$ that is $\leq_{\text{m}}^p$-complete for $\Sigma_k^p$, a set $\widehat{L}_{\Sigma_{k-1}^p}$ that is $\leq_{\text{m}}^p$-complete for $\Sigma_{k-1}^p$, and a set $L''_{\Sigma_{k-2}^p}$ that is $\leq_{\text{m}}^p$-complete for $\Sigma_{k-2}^p$ such that*

$$L'_{\Sigma_k^p} = \{x \mid (\exists y \in \Sigma^{|x|})(\forall z \in \Sigma^{|x|})[\langle x, y, z \rangle \in L''_{\Sigma_{k-2}^p}]\}$$

*and*

$$\widehat{L}_{\Sigma_{k-1}^p} = \{\langle x, y, z \rangle \mid |x| = |y| \land (\exists z')[(|x| = |zz'|) \land \langle x, y, zz' \rangle \notin L''_{\Sigma_{k-2}^p}]\}.$$

**Proof of Theorem 3.4** Let $m > 0$ and $k > 1$. Let $\widehat{L}_{\Sigma_{k-1}^p} \in \Sigma_{k-1}^p$ be as fixed in Notation 3.5, and let $L_{\Delta_{k-1}^p}$ and $L_{\text{DIFF}_m(\Sigma_k^p)}$ be any fixed $\leq_{\text{m}}^p$-complete sets for $\Delta_{k-1}^p$ and $\text{DIFF}_m(\Sigma_k^p)$, respectively; such languages exist, e.g., via the standard type of canonical-complete-set constructions involving padding and enumerations of clocked machines. From Lemma 3.2 it follows that $L_{\Delta_{k-1}^p} \tilde{\Delta} L_{\text{DIFF}_m(\Sigma_k^p)}$ is $\leq_{\text{m}}^p$-hard for $\Delta_{k-1}^p \boldsymbol{\Delta} \text{DIFF}_m(\Sigma_k^p)$. Since $\widehat{L}_{\Sigma_{k-1}^p} \tilde{\Delta} L_{\text{DIFF}_m(\Sigma_k^p)} \in \Sigma_{k-1}^p \boldsymbol{\Delta} \text{DIFF}_m(\Sigma_k^p)$ and by assumption $\Delta_{k-1}^p \boldsymbol{\Delta} \text{DIFF}_m(\Sigma_k^p) = \Sigma_{k-1}^p \boldsymbol{\Delta} \text{DIFF}_m(\Sigma_k^p)$, there exists a polynomial-time many-one reduction $h$ from $\widehat{L}_{\Sigma_{k-1}^p} \tilde{\Delta} L_{\text{DIFF}_m(\Sigma_k^p)}$ to $L_{\Delta_{k-1}^p} \tilde{\Delta} L_{\text{DIFF}_m(\Sigma_k^p)}$ (in light of the latter's $\leq_{\text{m}}^p$-hardness). So, for all $x_1, x_2, y_1, y_2 \in \Sigma^*$: if $h(\langle x_1, x_2 \rangle) = \langle y_1, y_2 \rangle$, then $((x_1 \in \widehat{L}_{\Sigma_{k-1}^p} \Leftrightarrow x_2 \notin L_{\text{DIFF}_m(\Sigma_k^p)})$ iff $(y_1 \in L_{\Delta_{k-1}^p} \Leftrightarrow y_2 \notin L_{\text{DIFF}_m(\Sigma_k^p)}))$. Equivalently, for all $x_1, x_2, y_1, y_2 \in \Sigma^*$: if $h(\langle x_1, x_2 \rangle) = \langle y_1, y_2 \rangle$, then

$$(x_1 \in \widehat{L}_{\Sigma_{k-1}^p} \Leftrightarrow x_2 \in L_{\text{DIFF}_m(\Sigma_k^p)}) \text{ if and only if } (y_1 \in L_{\Delta_{k-1}^p} \Leftrightarrow y_2 \in L_{\text{DIFF}_m(\Sigma_k^p)}). \qquad (3.1)$$

We can use $h$ to recognize some of $\overline{L_{\text{DIFF}_m(\Sigma_k^p)}}$ by a $\text{DIFF}_m(\Sigma_k^p)$ algorithm. In particular, we say that a string $x$ is *easy for length $n$* if there exists a string $x_1$ such that $|x_1| \leq n$ and $(x_1 \in \widehat{L}_{\Sigma_{k-1}^p} \Leftrightarrow y_1 \notin L_{\Delta_{k-1}^p})$ where $h(\langle x_1, x \rangle) = \langle y_1, y_2 \rangle$.

Let $p$ be a fixed polynomial, which will be exactly specified later in the proof. We have the following algorithm to test whether $x \in \overline{L_{\text{DIFF}_m(\Sigma_k^p)}}$ in the case that (our input) $x$ is an easy string for length $p(|x|)$. Guess $x_1$ with $|x_1| \leq p(|x|)$, let $h(\langle x_1, x \rangle) = \langle y_1, y_2 \rangle$, and accept if and only if $((x_1 \in \widehat{L}_{\Sigma_{k-1}^p} \Leftrightarrow y_1 \notin L_{\Delta_{k-1}^p}) \wedge y_2 \in L_{\text{DIFF}_m(\Sigma_k^p)})$.[2] This algorithm is not necessarily a $\text{DIFF}_m(\Sigma_k^p)$ algorithm, but it does inspire the following $\text{DIFF}_m(\Sigma_k^p)$ algorithm to test whether $x \in \overline{L_{\text{DIFF}_m(\Sigma_k^p)}}$ in the case that $x$ is an easy string for length $p(|x|)$.

Let $L_1, L_2, \ldots, L_m$ be languages in $\Sigma_k^p$ such that $L_{\text{DIFF}_m(\Sigma_k^p)} = L_1 - (L_2 - (L_3 - \cdots (L_{m-1} - L_m) \cdots))$ and $L_1 \supseteq L_2 \supseteq \cdots \supseteq L_{m-1} \supseteq L_m$ (this can be done, as this is simply the "telescoping" normal form of the levels of the boolean hierarchy over $\Sigma_k^p$, see [5,11,27]). For $1 \leq \ell \leq m$, define $L'_\ell$ as the language accepted by the following $\Sigma_k^p$ machine: On input $x$, guess $x_1$ with $|x_1| \leq p(|x|)$, let $h(\langle x_1, x \rangle) = \langle y_1, y_2 \rangle$, and accept if and only if $((x_1 \in \widehat{L}_{\Sigma_{k-1}^p} \Leftrightarrow y_1 \notin L_{\Delta_{k-1}^p}) \wedge y_2 \in L_\ell)$.

Note that $L'_\ell \in \Sigma_k^p$ for each $\ell$, and that $L'_1 \supseteq L'_2 \supseteq \cdots \supseteq L'_{m-1} \supseteq L'_m$. We will show that if $x$ is an easy string for length $p(|x|)$, then $x \in \overline{L_{\text{DIFF}_m(\Sigma_k^p)}}$ if and only if $x \in L'_1 - (L'_2 - \cdots (L'_{m-1} - L'_m) \cdots)$.

So suppose that $x$ is an easy string for length $p(|x|)$. Define $\ell'$ to be the unique integer such that (a) $0 \leq \ell' \leq m$, (b) $x \in L'_s$ for $1 \leq s \leq \ell'$, and (c) $x \notin L'_s$ for $s > \ell'$. It is immediate that $x \in L'_1 - (L'_2 - \cdots (L'_{m-1} - L'_m) \cdots)$ if and only if $\ell'$ is odd.

Let $w$ be some string such that $(\exists x_1 \in (\Sigma^*)^{\leq p(|x|)})(\exists y_1)[h(\langle x_1, x \rangle) = \langle y_1, w \rangle \wedge (x_1 \in \widehat{L}_{\Sigma_{k-1}^p} \Leftrightarrow y_1 \notin L_{\Delta_{k-1}^p})]$, and $w \in L_{\ell'}$ if $\ell' > 0$ (recall that $\ell'$ here is the $\ell'$ already set in the previous paragraph). Note that such a $w$ exists, since $x$ is easy for length $p(|x|)$ and by our definition of $\ell'$ it holds that $x \in L'_{\ell'}$. By the definition of $\ell'$ (namely, since $x \notin L'_s$ for $s > \ell'$), $w \notin L_s$ for all $s > \ell'$. It follows that $w \in L_{\text{DIFF}_m(\Sigma_k^p)}$ if and only if $\ell'$ is odd. It is clear, keeping in mind the definition of $h$, that $(x \in \overline{L_{\text{DIFF}_m(\Sigma_k^p)}} \Leftrightarrow w \in L_{\text{DIFF}_m(\Sigma_k^p)})$, $(w \in L_{\text{DIFF}_m(\Sigma_k^p)} \Leftrightarrow \ell'$ is odd$)$, and $(\ell'$ is odd $\Leftrightarrow x \in L'_1 - (L'_2 - \cdots (L'_{m-1} - L'_m) \cdots))$. So $x \in \overline{L_{\text{DIFF}_m(\Sigma_k^p)}} \Leftrightarrow x \in L'_1 - (L'_2 - \cdots (L'_{m-1} - L'_m) \cdots)$.

This completes the case where $x$ is easy, as $L'_1 - (L'_2 - \cdots (L'_{m-1} - L'_m) \cdots)$ in effect specifies a $\text{DIFF}_m(\Sigma_k^p)$ algorithm.

We say that $x$ is *hard for length $n$* if $|x| \leq n$ and $x$ is not easy for length $n$, i.e., if $|x| \leq n$ and for all $x_1$ with $|x_1| \leq n$, $(x_1 \in \widehat{L}_{\Sigma_{k-1}^p} \Leftrightarrow y_1 \in L_{\Delta_{k-1}^p})$, where $h(\langle x_1, x \rangle) = \langle y_1, y_2 \rangle$. Note that if $x$ is hard for $p(|x|)$, then $x \notin L'_1$.

---

[2]To understand what is going on here, simply note that if $(x_1 \in \widehat{L}_{\Sigma_{k-1}^p} \Leftrightarrow y_1 \notin L_{\Delta_{k-1}^p})$ holds, then by equation 3.1 we have $x \in \overline{L_{\text{DIFF}_m(\Sigma_k^p)}} \Leftrightarrow y_2 \in L_{\text{DIFF}_m(\Sigma_k^p)}$. Note also that both of $x_1 \in \widehat{L}_{\Sigma_{k-1}^p}$ and $y_1 \notin L_{\Delta_{k-1}^p}$ can be very easily tested by a machine that has a $\Sigma_{k-1}^p$ oracle.

If $x$ is a hard string for length $p(|x|)$, then $x$ induces a many-one reduction from $\left(\widehat{L}_{\Sigma_{k-1}^p}\right)^{\leq p(|x|)}$ to $L_{\Delta_{k-1}^p}$, namely, $\lambda x_1.f(x, x_1)$, where $f(x, x_1) = y_1$, where $y_1$ is the unique string such that $(\exists y_2)[h(\langle x_1, x\rangle) = \langle y_1, y_2\rangle]$. We will write $f_x$ for $\lambda x_1.f(x, x_1)$. Note that $f$ is computable in polynomial time.

So it is not hard to see that if we choose $p$ appropriately large, then a hard string $x$ for length $p(|x|)$ induces $\Sigma_{k-1}^p$ algorithms for $(L_1)^{=|x|}, (L_2)^{=|x|}, \ldots, (L_m)^{=|x|}$ (essentially since each is in $\Sigma_k^p = \mathrm{NP}^{\Sigma_{k-1}^p}$, $\widehat{L}_{\Sigma_{k-1}^p}$ is $\leq_\mathrm{m}^p$-complete for $\Sigma_{k-1}^p$, and $\mathrm{NP}^{\Delta_{k-1}^p} = \Sigma_{k-1}^p$), which we can use to obtain a $\mathrm{DIFF}_m(\Sigma_{k-1}^p)$ algorithm for $\left(L_{\mathrm{DIFF}_m(\Sigma_k^p)}\right)^{=|x|}$, and thus certainly a $\mathrm{DIFF}_m(\Sigma_k^p)$ algorithm for $\left(\overline{L_{\mathrm{DIFF}_m(\Sigma_k^p)}}\right)^{=|x|}$.

However, there is a problem. The problem is that we cannot combine the $\mathrm{DIFF}_m(\Sigma_k^p)$ algorithms for easy and hard strings into one $\mathrm{DIFF}_m(\Sigma_k^p)$ algorithm for $\overline{L_{\mathrm{DIFF}_m(\Sigma_k^p)}}$ that works all strings. Why? It is too difficult to decide whether a string is easy or hard; to decide this deterministically takes one query to $\Sigma_k^p$, and we cannot do that in a $\mathrm{DIFF}_m(\Sigma_k^p)$ algorithm. This is also the reason why the methods from [13] failed to prove that if $\mathrm{P}\mathbf{\Delta}\Sigma_2^p = \mathrm{NP}\mathbf{\Delta}\Sigma_2^p$, then $\Sigma_2^p = \Pi_2^p$. Recall from the introduction that the latter theorem was proven by Buhrman and Fortnow [4]. We will generalize their technique at this point. In particular, the following lemma, which we will prove after we have finished the proof of this theorem, establishes a generalized version of the technique from [4]. It has been extended to deal with arbitrary levels of the polynomial hierarchy and to be useful in settings involving boolean hierarchies.

**Lemma 3.6** *Let $k > 1$. For all $L \in \Sigma_k^p$, there exist a polynomial $q$ and a set $\widehat{L} \in \Pi_{k-1}^p$ such that*

1. *for each natural number $n'$, $q(n') \geq n'$,*

2. *$\widehat{L} \subseteq \overline{L}$, and*

3. *if $x$ is hard for length $q(|x|)$, then $(x \in \overline{L} \Leftrightarrow x \in \widehat{L})$.*

We defer the proof of Lemma 3.6 and first finish the current proof. From Lemma 3.6, it follows that there exist sets $\widehat{L_1}, \widehat{L_2}, \ldots, \widehat{L_m} \in \Pi_{k-1}^p$ and polynomials $q_1, q_2, \ldots, q_m$ with the following properties for all $1 \leq \ell \leq m$:

1. $\widehat{L_\ell} \subseteq \overline{L_\ell}$, and

2. if $x$ is hard for length $q_\ell(|x|)$, then $(x \in \overline{L_\ell} \Leftrightarrow x \in \widehat{L_\ell})$.

Take $p$ to be an (easy-to-compute—we may without loss of generality require that there is an $t$ such that it is of the form $n^t + t$) polynomial such that $p$ is at least as large as all the $q_\ell$s, i.e., such that, for each natural number $n'$, we have $p(n') \geq \max\{q_1(n'), \ldots, q_m(n')\}$. By the definition of hardness and condition 1 of Lemma 3.6, if $x$ is hard for length $p(|x|)$, then $x$ is hard for length $q_\ell(|x|)$ for all $1 \leq \ell \leq m$. As promised earlier, we have now specified $p$. Define $\widehat{L}_{\text{DIFF}_m(\Sigma_k^p)}$ as follows: On input $x$, guess $\ell$, $\ell$ even, $0 \leq \ell \leq m$, and accept if and only if both (a) $x \in L_\ell$ or $\ell = 0$, and (b) if $\ell < m$, then $x \in \widehat{L_{\ell+1}}$. Clearly, $\widehat{L}_{\text{DIFF}_m(\Sigma_k^p)} \in \Sigma_k^p$. In addition, this set inherits certain properties from the $\widehat{L}_\ell$s. In particular, in light of the definition of $\widehat{L}_{\text{DIFF}_m(\Sigma_k^p)}$, the definitions of the $\widehat{L}_\ell$s, and the fact that:

$$x \in \overline{L_{\text{DIFF}_m(\Sigma_k^p)}} \Leftrightarrow (\exists \ell, 0 \leq \ell \leq m, \ell \text{ even})[(\ell \neq 0 \Rightarrow x \in L_\ell) \wedge (\ell \neq m \Rightarrow x \in \overline{L_{\ell+1}})],$$

we have that the following properties hold:

1. $\widehat{L}_{\text{DIFF}_m(\Sigma_k^p)} \subseteq \overline{L_{\text{DIFF}_m(\Sigma_k^p)}}$, and

2. if $x$ is hard for length $p(|x|)$, then $(x \in \overline{L_{\text{DIFF}_m(\Sigma_k^p)}} \Leftrightarrow x \in \widehat{L}_{\text{DIFF}_m(\Sigma_k^p)})$.

Finally, we are ready to give the algorithm. Recall that $L_1', L_2', \ldots, L_m'$ are sets in $\Sigma_k^p$ such that: (1) $L_1' \supseteq L_2' \supseteq \cdots \supseteq L_{m-1}' \supseteq L_m'$, and (2) if $x$ is easy for length $p(|x|)$, then $x \in \overline{L_{\text{DIFF}_m(\Sigma_k^p)}}$ if and only if $x \in L_1' - (L_2' - (L_3' - \cdots (L_{m-1}' - L_m') \cdots))$, and (3) if $x$ is hard for length $p(|x|)$, then $x \notin L_1'$. We claim that for all $x$, $(x \in \overline{L_{\text{DIFF}_m(\Sigma_k^p)}} \Leftrightarrow x \in (L_1' \cup \widehat{L}_{\text{DIFF}_m(\Sigma_k^p)}) - (L_2' - (L_3' - \cdots (L_{m-1}' - L_m') \cdots)))$, which completes the proof of Theorem 3.4, as $\Sigma_k^p$ is closed under union.

($\Rightarrow$): Let $x \in \overline{L_{\text{DIFF}_m(\Sigma_k^p)}}$. If $x$ is easy for length $p(|x|)$, then $x \in L_1' - (L_2' - (L_3' - \cdots (L_{m-1}' - L_m') \cdots))$, and so certainly $x \in (L_1' \cup \widehat{L}_{\text{DIFF}_m(\Sigma_k^p)}) - (L_2' - (L_3' - \cdots (L_{m-1}' - L_m') \cdots))$. If $x$ is hard for length $p(|x|)$, then $x \in \widehat{L}_{\text{DIFF}_m(\Sigma_k^p)}$ and $x \notin L_\ell'$ for all $\ell$ (since $x \notin L_1'$ and $L_1' \supseteq L_2' \supseteq \cdots \supseteq L_m'$). Thus, $x \in (L_1' \cup \widehat{L}_{\text{DIFF}_m(\Sigma_k^p)}) - (L_2' - (L_3' - \cdots (L_{m-1}' - L_m') \cdots))$.

($\Leftarrow$): Suppose $x \in (L_1' \cup \widehat{L}_{\text{DIFF}_m(\Sigma_k^p)}) - (L_2' - (L_3' - \cdots (L_{m-1}' - L_m') \cdots))$. If $x \in \widehat{L}_{\text{DIFF}_m(\Sigma_k^p)}$, then $x \in \overline{L_{\text{DIFF}_m(\Sigma_k^p)}}$. If $x \notin \widehat{L}_{\text{DIFF}_m(\Sigma_k^p)}$, then $x \in L_1' - (L_2' - (L_3' - \cdots (L_{m-1}' - L_m') \cdots))$ and so $x$ must be easy for length $p(|x|)$ (as $x \in L_1'$, and this is possible only if $x$ is easy for length $p(|x|)$). However, this says that $x \in \overline{L_{\text{DIFF}_m(\Sigma_k^p)}}$. ∎

Having completed the proof of Theorem 3.4, we now return to the deferred proof of the lemma used within that theorem's proof.

**Proof of Lemma 3.6.** Let $L \in \Sigma_k^p$. We need to show that there exist a polynomial $q$ and a set $\widehat{L} \in \Pi_{k-1}^p$ such that

1. for each natural number $n'$, $q(n') \geq n'$,

2. $\widehat{L} \subseteq \overline{L}$, and

3. if $x$ is hard for length $q(|x|)$, then $(x \in \overline{L} \Leftrightarrow x \in \widehat{L})$.

From Notation 3.5, we know that $L'_{\Sigma^p_k}$ is $\leq^p_m$-complete for $\Sigma^p_k$, $\widehat{L}_{\Sigma^p_{k-1}} \in \Sigma^p_{k-1}$, $L''_{\Sigma^p_{k-2}} \in \Sigma^p_{k-2}$, and

1. $L'_{\Sigma^p_k} = \{x \mid (\exists y \in \Sigma^{|x|})(\forall z \in \Sigma^{|x|})[\langle x, y, z \rangle \in L''_{\Sigma^p_{k-2}}]\}$, and

2. $\widehat{L}_{\Sigma^p_{k-1}} = \{\langle x, y, z \rangle \mid |x| = |y| \wedge (\exists z')[(|x| = |zz'|) \wedge \langle x, y, zz' \rangle \notin L''_{\Sigma^p_{k-2}}]\}$.

Note that $\overline{L'_{\Sigma^p_k}} = \{x \mid (\forall y \in \Sigma^{|x|})(\exists z \in \Sigma^{|x|})[\langle x, y, z \rangle \notin L''_{\Sigma^p_{k-2}}]\}$.

Since $L \in \Sigma^p_k$, and $L'_{\Sigma^p_k}$ is $\leq^p_m$-complete for $\Sigma^p_k$, there exists a polynomial-time computable function $g$ such that, for all $x$, $(x \in L \Leftrightarrow g(x) \in L'_{\Sigma^p_k})$.

Let $q$ be such that (a) $(\forall x \in \Sigma^n)(\forall y \in \Sigma^{|g(x)|})(\forall z \in (\Sigma^*)^{\leq |g(x)|})[q(n) \geq |\langle g(x), y, z \rangle|]$ and (b) $(\forall \widehat{m} \geq 0)[q(\widehat{m}+1) > q(\widehat{m}) > 0]$. Note that we have ensured that for each natural number $n'$, $q(n') \geq n'$.

If $x$ is a hard string for length $p(|x|)$, then $x$ induces a many-one reduction from $\left(\widehat{L}_{\Sigma^p_{k-1}}\right)^{\leq p(|x|)}$ to $L_{\Delta^p_{k-1}}$, namely, $\lambda x_1.f(x, x_1)$, where $f(x, x_1) = y_1$, where $y_1$ is the unique string such that $(\exists y_2)[h(\langle x_1, x \rangle) = \langle y_1, y_2 \rangle]$. (This is the $h$ from the proof of Theorem 3.4. One should treat the current proof as if it occurs immediately after the statement of Lemma 3.6.) We will write $f_x$ for $\lambda x_1.f(x, x_1)$. Note that $f$ is computable in polynomial time.

Let $\widehat{L}$ be the language accepted by the following $\Pi^p_{k-1}$ machine:[3]

On input $x$:
>    Compute $g(x)$
>    Guess $y$ such that $|y| = |g(x)|$
>    Set $w = \epsilon$ (i.e., the empty string)
>    While $|w| < |g(x)|$
>    >    if the $\Delta^p_{k-1}$ algorithm for $\widehat{L}_{\Sigma^p_{k-1}}$ induced by $x$ accepts $\langle g(x), y, w0 \rangle$
>    >    (that is, if $f_x(\langle g(x), y, w0 \rangle) \in L_{\Delta^p_{k-1}}$),
>    >    then $w = w0$
>    >    else $w = w1$
>    Accept if and only if $\langle g(x), y, w \rangle \notin L''_{\Sigma^p_{k-2}}$.

It remains to show that $\widehat{L}$ thus defined fulfills the properties of Lemma 3.6. First note that the machine described above is clearly a $\Pi^p_{k-1}$ machine. To show that $\widehat{L} \subseteq \overline{L}$, suppose that $x \in \widehat{L}$. Then (keeping in mind the comments of footnote 3) for every $y \in \Sigma^{|g(x)|}$, there exists a string $w \in \Sigma^{|g(x)|}$ such that $\langle g(x), y, w \rangle \notin L''_{\Sigma^p_{k-2}}$. This implies that $g(x) \in \overline{L'_{\Sigma^p_k}}$, and thus that $x \in \overline{L}$.

---

[3] For $k > 1$, $\Pi^p_{k-1} = \mathrm{coNP}^{\Sigma^p_{k-2}}$, and by a $\Pi^p_{k-1}$ machine we mean, for the duration of this proof, a co-nondeterministic machine with a $\Sigma^p_{k-2}$ oracle. A co-nondeterministic machine by definition accepts iff *all* of its computation paths are accepting paths.

Finally, suppose that $x$ is hard for length $q(|x|)$ and that $x \in \overline{L}$. We have to show that $x \in \widehat{L}$. Since $x \in \overline{L}$, $g(x) \in \overline{L'_{\Sigma^p_k}}$. So, $(\forall y \in \Sigma^{|g(x)|})(\exists z \in \Sigma^{|g(x)|})[\langle g(x), y, z\rangle \notin L''_{\Sigma^p_{k-2}}]$. Since $x$ is hard for length $q(|x|)$, $(\forall y \in \Sigma^{|g(x)|})(\forall w \in (\Sigma^*)^{\leq |g(x)|})[\langle g(x), y, w\rangle \in \widehat{L}_{\Sigma^p_{k-1}} \Leftrightarrow f_x(\langle g(x), y, w\rangle) \in L_{\Delta^p_{k-1}}]$. It follows that the algorithm above will find, for every $y \in \Sigma^{|g(x)|}$, a witness $w$ such that $\langle g(x), y, w\rangle \notin L''_{\Sigma^p_{k-2}}$, and thus the algorithm will accept $x$. ∎

## 4 Downward Collapse from Closure Under Complementation

Recall that the $s = 1$ case of this section's main result, Theorem 4.1, is used along with Theorem 3.4 to establish Theorem 3.3. However, Theorem 4.1 is of interest in its own right as a reflection of how closure under complementation of even quite general symmetric difference classes implies a downward collapse. Selivanov [22,23] shows that if certain symmetric difference classes are closed under complementation, then the polynomial hierarchy collapses. His result is, however, very different than this section's main result, Theorem 4.1, as Selivanov collapses the polynomial hierarchy to a higher level, and thus shows merely an upward translation of equality. In contrast, our Theorem 4.1 collapses the difference hierarchy over $\Sigma^p_k$ to a level that is contained in the classes of its complementation hypothesis—thus obtaining a *downward* translation of equality. Also, we note that Theorem 4.1 implies a collapse of the polynomial hierarchy to a class a full level lower in the difference hierarchy over $\Sigma^p_{k+1}$ than could be concluded without our downward collapse result (namely to $\text{DIFF}_m(\Sigma^p_k)\boldsymbol{\Delta}\text{DIFF}_{m-1}(\Sigma^p_{k+1})$, in light of the strongest known "BH/PH-collapse connection," see [12,20] and the related discussion in Section 5).

**Theorem 4.1** *Let $s, m > 0$ and $0 < i < k - 1$. If $\text{DIFF}_s(\Sigma^p_i)\boldsymbol{\Delta}\text{DIFF}_m(\Sigma^p_k)$ is closed under complementation, then $\text{DIFF}_m(\Sigma^p_k) = \text{coDIFF}_m(\Sigma^p_k)$.*

Before proving Theorem 4.1, we fix some useful sets and the notation we will use for them.[4]

---

[4]The reason these sets exist is similar to the reason that the sets of Notation 3.5 exist, but may at first seem a bit confusing, due to the fact that in Notation 4.2 $\widetilde{L}_{\Sigma^p_{i+1}}$ is being treated as a set of strings in its own definition but as a set of pairs of strings in the definition of $L^\dagger_{\Sigma^p_{i+2}}$. However, since the pairing function maps strings to strings, this isn't a problem; it merely requires some pairing in forming the sets. For example, to give the intuition of what is going on, consider the sets:

- $A_{\Sigma^p_0} = \{\langle\langle F, v\rangle, w\rangle \,\big|\, v$ specifies assignments to the first half of $F$'s variables and $w$ specifies assignments to the second half of $F$'s variables and $F$ under the (complete) assignment specified by $v$ and $w$ evaluates to false$\}$.

**Notation 4.2** *For each $i \geq 1$, fix three sets $L_{\Sigma_i^p}$, $\widetilde{L}_{\Sigma_{i+1}^p}$, and $L_{\Sigma_{i+2}^p}^{\dagger}$ that are $\leq_{\mathrm{m}}^p$-complete for $\Sigma_i^p$, $\Sigma_{i+1}^p$, and $\Sigma_{i+2}^p$, respectively, and that satisfy*

$$\widetilde{L}_{\Sigma_{i+1}^p} = \{x \mid (\exists y \in \Sigma^{|x|})[\langle x, y\rangle \notin L_{\Sigma_i^p}]\},$$

*and*

$$L_{\Sigma_{i+2}^p}^{\dagger} = \{x \mid (\exists y \in \Sigma^{|x|})[\langle x, y\rangle \notin \widetilde{L}_{\Sigma_{i+1}^p}]\}.$$

For each $i \geq 1$, let $L_{\Pi_i^p} = \overline{L_{\Sigma_i^p}}$ and define $L_{\mathrm{DIFF}_1(\Pi_i^p)} = L_{\Pi_i^p}$ and for all $j \geq 2$, $L_{\mathrm{DIFF}_j(\Pi_i^p)} = \{\langle x, y\rangle \mid x \in L_{\Pi_i^p} \wedge y \notin L_{\mathrm{DIFF}_{j-1}(\Pi_i^p)}\}$. It is not hard to see that $L_{\mathrm{DIFF}_j(\Pi_i^p)}$ is many-one complete for $\mathrm{DIFF}_j(\Pi_i^p)$ for all $j \geq 1$. Note also that $\mathrm{DIFF}_j(\Pi_i^p) = \mathrm{DIFF}_j(\Sigma_i^p)$ if $j$ is even and $\mathrm{DIFF}_j(\Pi_i^p) = \mathrm{coDIFF}_j(\Sigma_i^p)$ if $j$ is odd. Let $L_{\mathrm{DIFF}_s(\Sigma_i^p)} = L_{\mathrm{DIFF}_s(\Pi_i^p)}$ if $s$ is even and $L_{\mathrm{DIFF}_s(\Sigma_i^p)} = \overline{L_{\mathrm{DIFF}_s(\Pi_i^p)}}$ if $s$ is odd. Then $L_{\mathrm{DIFF}_s(\Sigma_i^p)}$ is $\leq_{\mathrm{m}}^p$-complete for $\mathrm{DIFF}_s(\Sigma_i^p)$.

**Proof of Theorem 4.1** Let $s, m > 0$ and $0 < i < k-1$. $L_{\mathrm{DIFF}_s(\Sigma_i^p)} \tilde{\Delta} L_{\mathrm{DIFF}_m(\Sigma_k^p)}$ is $\leq_{\mathrm{m}}^p$-hard for $\mathrm{DIFF}_s(\Sigma_i^p) \mathbf{\Delta} \mathrm{DIFF}_m(\Sigma_k^p)$ by Lemma 3.2 and $L_{\mathrm{DIFF}_s(\Sigma_i^p)} \tilde{\Delta} L_{\mathrm{DIFF}_m(\Sigma_k^p)}$ is clearly in $\mathrm{DIFF}_s(\Sigma_i^p) \mathbf{\Delta} \mathrm{DIFF}_m(\Sigma_k^p)$. Since by assumption $\mathrm{DIFF}_s(\Sigma_i^p) \mathbf{\Delta} \mathrm{DIFF}_m(\Sigma_k^p)$ is closed under complementation, there exists a polynomial-time many-one reduction $h$ from $L_{\mathrm{DIFF}_s(\Sigma_i^p)} \tilde{\Delta} L_{\mathrm{DIFF}_m(\Sigma_k^p)}$ to its complement. That is, for all $x_1, x_2, y_1, y_2 \in \Sigma^*$ it holds that: if $h(\langle x_1, x_2\rangle) = \langle y_1, y_2\rangle$, then $(\langle x_1, x_2\rangle \in L_{\mathrm{DIFF}_s(\Sigma_i^p)} \tilde{\Delta} L_{\mathrm{DIFF}_m(\Sigma_k^p)} \Leftrightarrow \langle y_1, y_2\rangle \notin L_{\mathrm{DIFF}_s(\Sigma_i^p)} \tilde{\Delta} L_{\mathrm{DIFF}_m(\Sigma_k^p)})$. Equivalently, for all $x_1, x_2, y_1, y_2 \in \Sigma^*$:

**Fact 1:**
if $h(\langle x_1, x_2\rangle) = \langle y_1, y_2\rangle$, then:

$(x_1 \in L_{\mathrm{DIFF}_s(\Sigma_i^p)} \Leftrightarrow x_2 \notin L_{\mathrm{DIFF}_m(\Sigma_k^p)})$ if and only if

$$(y_1 \in L_{\mathrm{DIFF}_s(\Sigma_i^p)} \Leftrightarrow y_2 \in L_{\mathrm{DIFF}_m(\Sigma_k^p)}).$$

We can use $h$ to recognize some of $\overline{L_{\mathrm{DIFF}_m(\Sigma_k^p)}}$ by a $\mathrm{DIFF}_m(\Sigma_k^p)$ algorithm. In particular, we say that a string $x$ is *easy for length $n$* if there exists a string $x_1$ such that $|x_1| \leq n$ and $(x_1 \in L_{\mathrm{DIFF}_s(\Sigma_i^p)} \Leftrightarrow y_1 \in L_{\mathrm{DIFF}_s(\Sigma_i^p)})$ where $h(\langle x_1, x\rangle) = \langle y_1, y_2\rangle$.

- $A_{\Sigma_1^p} = \{\langle F, v\rangle \mid v$ specifies assignments to the first half of $F$'s variables and there exists a $w$ specifying assignments to the second half of $F$'s variables such that $\langle\langle F, v\rangle, w\rangle \notin A_{\Sigma_0^p}\}$.
- $A_{\Sigma_2^p} = \{F \mid$ there exists a $v$ specifying assignments to the first half of $F$'s variables such that $\langle F, v\rangle \notin A_{\Sigma_1^p}\}$.

These sets are easily seen to be respectively $\leq_{\mathrm{m}}^p$-complete for $\Sigma_0^p$ (trivial, in this case), $\Sigma_1^p$, and $\Sigma_2^p$. Caveat: These are not exactly the sets of Notation 4.2, as here we have not been careful to make sure the quantified lengths are exactly the same length as the input, and we have been sloppy about what "half" means when the number of variables is odd; however, this example should make it clearer that sets satisfying Notation 4.2 exist.

Let $p$ be a fixed polynomial, which will be exactly specified later in the proof. We have the following algorithm to test whether $x \in \overline{L_{\mathrm{DIFF}_m(\Sigma_k^p)}}$ in the case that (our input) $x$ is an easy string for length $p(|x|)$. On input $x$, guess $x_1$ with $|x_1| \leq p(|x|)$, let $h(\langle x_1, x \rangle) = \langle y_1, y_2 \rangle$, and accept if and only if $((x_1 \in L_{\mathrm{DIFF}_s(\Sigma_i^p)} \Leftrightarrow y_1 \in L_{\mathrm{DIFF}_s(\Sigma_i^p)}) \wedge y_2 \in L_{\mathrm{DIFF}_m(\Sigma_k^p)})$. This algorithm is not necessarily a $\mathrm{DIFF}_m(\Sigma_k^p)$ algorithm, but in the same way as in the proof of Theorem 3.4, we can construct sets $L_1', L_2', \ldots, L_m' \in \Sigma_k^p$ such that if $x$ is an easy string for length $p(|x|)$, then $x \in \overline{L_{\mathrm{DIFF}_m(\Sigma_k^p)}}$ if and only if $x \in L_1' - (L_2' - (L_3' - \cdots (L_{m-1}' - L_m') \cdots))$.

We say that $x$ is *hard for length* $n$ if $|x| \leq n$ and $x$ is not easy for length $n$, i.e., if $|x| \leq n$ and, for all $x_1$ with $|x_1| \leq n$, $(x_1 \in L_{\mathrm{DIFF}_s(\Sigma_i^p)} \Leftrightarrow y_1 \notin L_{\mathrm{DIFF}_s(\Sigma_i^p)})$, where $h(\langle x_1, x \rangle) = \langle y_1, y_2 \rangle$.

If $x$ is a hard string for length $n$, then $x$ induces a many-one reduction from $\left(L_{\mathrm{DIFF}_s(\Sigma_i^p)}\right)^{\leq n}$ to $\overline{L_{\mathrm{DIFF}_s(\Sigma_i^p)}}$, namely, $\lambda x_1 . f(x, x_1)$, where $f(x, x_1) = y_1$, where $y_1$ is the unique string such that $(\exists y_2)[h(\langle x_1, x \rangle) = \langle y_1, y_2 \rangle]$. We will write $f_x$ for $\lambda x_1 . f(x, x_1)$. Note that $f$ is computable in polynomial time.

It is known that a collapse of the boolean hierarchy over $\Sigma_i^p$ implies a collapse of the polynomial hierarchy. A long series of papers studied the question to what level the polynomial hierarchy collapses in that case. The best known results (e.g., [8,3,13,20,12], see especially the strongest such connection, which is that obtained independently in [20] and [12]) conclude a collapse of the polynomial hierarchy to a level within the boolean hierarchy over $\Sigma_{i+1}^p$. Though a hard string for length $n$ only induces a many-one reduction between initial segments of $L_{\mathrm{DIFF}_s(\Sigma_i^p)}$ and $\overline{L_{\mathrm{DIFF}_s(\Sigma_i^p)}}$, we would nevertheless like to derive at least a $\mathrm{P}^{\Sigma_{k-1}^p}$ algorithm for some of $L_{\Sigma_{i+2}^p}^\dagger$. The following lemma does exactly that.

**Lemma 4.3** *Let $s, m > 0$, and $0 < i < k - 1$, and suppose that $\mathrm{DIFF}_s(\Sigma_i^p) \boldsymbol{\Delta} \mathrm{DIFF}_m(\Sigma_k^p) = \mathrm{co}(\mathrm{DIFF}_s(\Sigma_i^p) \boldsymbol{\Delta} \mathrm{DIFF}_m(\Sigma_k^p))$. There exist a set $D \in \mathrm{P}^{\Sigma_{i+1}^p}$ and a polynomial $r$ such that for all $n$, (a) $r(n+1) > r(n) > 0$ and (b) for all $x \in \Sigma^*$, if $x$ is a hard string for length $r(n)$ then for all $y \in (\Sigma^*)^{\leq n}$,*
$$y \in L_{\Sigma_{i+2}^p}^\dagger \Leftrightarrow \langle x, 1^n, y \rangle \in D.$$

We defer the proof of Lemma 4.3 and first finish the proof of the current theorem.

We will use the result of Lemma 4.3 to obtain a $\mathrm{P}^{\Sigma_{k-1}^p}$ algorithm that for any string $x$ that is hard for length $p(|x|)$ will determine whether $x \in L_{\mathrm{DIFF}_m(\Sigma_k^p)}$.

Let $L_1, L_2, \ldots, L_m$ be languages in $\Sigma_k^p$ such that $L_{\mathrm{DIFF}_m(\Sigma_k^p)} = L_1 - (L_2 - (L_3 - \cdots (L_{m-1} - L_m) \cdots))$. Since $L_{\Sigma_k^p}$ is complete for $\Sigma_k^p$, there exist functions $g_1, \cdots, g_m$ that many-one reduce $L_1, \cdots, L_m$ to $L_{\Sigma_k^p}$, respectively. Let the output sizes of all the $g_j$'s be bounded by the polynomial $p'$, which without loss of generality satisfies $(\forall \widehat{m} \geq 0)[p'(\widehat{m}+1) > p'(\widehat{m}) > 0]$. So there exists a polynomial-time machine that queries strings of length at most $p'(n)$ on inputs of length $n$ to $L_{\Sigma_k^p}$ and that accepts $L_{\text{DIFF}_m(\Sigma_k^p)}$.

A $\Sigma_0^p$ machine is an oracle P machine; for $z \geq 1$, a $\Sigma_z^p$ machine is a polynomial-time-bounded $z$-alternation-block-bounded oracle machine with the first alteration block existential. For example, the class of languages accepted by $\Sigma_2^p$ machines allowed $\Sigma_3^p$ oracles is $\Sigma_5^p$. Let $M$ be a $\Sigma_{k-(i+2)}^p$ machine recognizing $L_{\Sigma_k^p}$ with oracle queries to $L_{\Sigma_{i+2}^p}^\dagger$ and running in time $q'$ for some polynomial $q'$ satisfying $(\forall \widehat{m} \geq 0)[q'(\widehat{m}+1) > q'(\widehat{m}) > 0]$. Let $p$ be an easily computable polynomial satisfying $(\forall \widehat{m} \geq 0)[p(\widehat{m}+1) > p(\widehat{m}) > 0]$ and for all $n$, $p(n) \geq r(q'(p'(n)))$, where $r$ is the polynomial of Lemma 4.3. As promised, we now have specified $p$.

If $x$ is a hard string for length $p(|x|)$, then $x$ is also a hard string for length $r(q'(p'(|x|)))$. So, by Lemma 4.3, for all $y \in (\Sigma^*)^{\leq q'(p'(|x|))}$, $y \in L_{\Sigma_{i+2}^p}^\dagger \Leftrightarrow \langle x, 1^n, y \rangle \in D$. Define the following $P^{\Sigma_{k-1}^p}$ set $E$: on input $\langle x, z \rangle$, simulate $M$ on input $z$ and replace every query $y$ to $L_{\Sigma_{i+2}^p}^\dagger$ by query $\langle x, 1^n, y \rangle$ to $D$. (Note that if $i < k-2$, $E$ is even in $\Sigma_{k-1}^p$.) Clearly, for all $x \in \Sigma^*$, if $x$ is a hard string for length $p(|x|)$ then for all $z \in (\Sigma^*)^{\leq p'(|x|)}$, $\langle x, z \rangle \in E$ if and only if $z \in L_{\Sigma_k^p}$.

Recall that there exists a polynomial-time machine that determines whether $x \in L_{\text{DIFF}_m(\Sigma_k^p)}$ with oracle queries of length at most $p'(|x|)$ to $L_{\Sigma_k^p}$. If $x$ is a hard string for length $p(|x|)$, we can replace every query $z$ to $L_{\Sigma_k^p}$ by query $\langle x, z \rangle$ to $E$. We have now defined a $P^{P^{\Sigma_{k-1}^p}} = P^{\Sigma_{k-1}^p}$ algorithm that for any string $x$ that is hard for length $p(|x|)$ will determine whether $x \in L_{\text{DIFF}_m(\Sigma_k^p)}$.

However, now we have an outright $\text{DIFF}_m(\Sigma_k^p)$ algorithm for $\overline{L_{\text{DIFF}_m(\Sigma_k^p)}}$: For $1 \leq \ell \leq m$ define an $\text{NP}^{\Sigma_{k-1}^p}$ machine $N_\ell$ as follows: On input $x$, the NP base machine of $N_\ell$ executes the following algorithm:

1. Using its $\Sigma_{k-1}^p$ oracle, it deterministically determines whether the input $x$ is an easy string for length $p(|x|)$. This can be done, as checking whether the input is an easy string for length $p(|x|)$ can be done by one query to $\Sigma_{i+1}^p$, and $i+1 \leq k-1$ by our $i < k-1$ hypothesis.

2. If the previous step determined that the input is not an easy string, then the input must be a hard string for length $p(|x|)$. If $\ell = 1$, then simulate the $P^{\Sigma_{k-1}^p}$ algorithm for hard strings to determine whether $x \in L_{\text{DIFF}_m(\Sigma_k^p)}$ and accept if and only if $x \notin L_{\text{DIFF}_m(\Sigma_k^p)}$. If $\ell > 1$, then reject.

3. If the first step determined that the input $x$ is easy for length $p(|x|)$, then our NP machine simulates (using itself and its oracle) the $\Sigma_k^p$ algorithm for $L'_\ell$ on input $x$.

Note that the $\Sigma_{k-1}^p$ oracle in the above algorithm is being used for a number of different sets. However, as $\Sigma_{k-1}^p$ is closed under disjoint union, this presents no problem as we can use the disjoint union of the sets, while modifying the queries so they address the appropriate part of the disjoint union.

It follows that, for all $x$, $x \in \overline{L_{\mathrm{DIFF}_m(\Sigma_k^p)}}$ if and only if $x \in L(N_1) - (L(N_2) - (L(N_3) - \cdots (L(N_{m-1}) - L(N_m)) \cdots ))$. Since $\overline{L_{\mathrm{DIFF}_m(\Sigma_k^p)}}$ is complete for $\mathrm{coDIFF}_m(\Sigma_k^p)$, it follows that $\mathrm{DIFF}_m(\Sigma_k^p) = \mathrm{coDIFF}_m(\Sigma_k^p)$. ∎

We now give the proof of Lemma 4.3. This proof should be seen in the context of the proof of Theorem 4.1 and Notation 4.2 as some notations we are going to use are defined there.

**Proof of Lemma 4.3** Our proof generalizes a proof from [3]. Let $\langle \cdots \rangle$ be a pairing function that maps sequences of up to $2s+2$ of strings over $\Sigma^*$ to $\Sigma^*$ having the standard properties such as polynomial-time computability and invertibility, etc. Let $t$ be a polynomial such that $|\langle x_1, x_2, \ldots, x_j \rangle| \leq t(\max\{|x_1|, |x_2|, \ldots, |x_j|\})$ for all $1 \leq j \leq 2s+2$ and all $x_1, x_2, \ldots, x_j \in \Sigma^*$. Without loss of generality let $t$ be such that $t(n+1) > t(n) > 0$ for all $n$. Define $t^{(0)}(n) = n$ and $t^{(j)}(n) = \underbrace{t(t(\cdots t(n) \cdots))}_{j \text{ times}}$ for all $n$ and all $j \geq 1$.

Define $r'$ to be a polynomial such that $r'(n+1) > r'(n) > 0$ and $r'(n) \geq t^{(s-1)}(n)$ for all $n$. Let $n$ be an integer. Suppose that $x$ is a hard string for length $r'(n)$, where hardness is defined as in the proof of Theorem 4.1. Then (recall the sets fixed/named in Notation 4.2), for all $y$ such that $|y| \leq r'(n)$,

$$y \in L_{\mathrm{DIFF}_s(\Sigma_i^p)} \Leftrightarrow f_x(y) \notin L_{\mathrm{DIFF}_s(\Sigma_i^p)},$$

or equivalently

$$y \in L_{\mathrm{DIFF}_s(\Pi_i^p)} \Leftrightarrow f_x(y) \notin L_{\mathrm{DIFF}_s(\Pi_i^p)}.$$

Recall that $f_x = \lambda y.f(x, y)$ and that $f$ can be computed in polynomial time. If $s > 1$, let $y = \langle y_1, y_2 \rangle$ and let $f_x(y) = \langle z_1, z_2 \rangle$. Then, for all $y_1, y_2 \in \Sigma^*$ such that $|y_1| \leq n$ and $|y_2| \leq t^{(s-2)}(n)$,

$$y_1 \in L_{\Pi_i^p} \wedge y_2 \notin L_{\mathrm{DIFF}_{s-1}(\Pi_i^p)} \Leftrightarrow z_1 \notin L_{\Pi_i^p} \vee z_2 \in L_{\mathrm{DIFF}_{s-1}(\Pi_i^p)}. \tag{4.2}$$

If $s > 1$, we say that $y_1$ is $s$-easy for length $n$ if and only if $|y_1| \leq n$ and $(\exists y_2 \, |y_2| \leq t^{(s-2)}(n))[z_1 \notin L_{\Pi_i^p}]$. $y_1$ is said to be $s$-hard for length $n$ if and only if $|y_1| \leq n$, $y_1 \in L_{\Pi_i^p}$, and $(\forall y_2 \, |y_2| \leq t^{(s-2)}(n))[z_1 \in L_{\Pi_i^p}]$. Observe that the above notions are defined with respect to our hard string $x$, since $z_1$ depends on $x$, $y_1$, and $y_2$. Furthermore, according to equation 4.2, if $y_1$ is $s$-easy for length $n$ then $y_1 \in L_{\Pi_i^p}$.

Suppose there exists an $s$-hard string $\omega_s$ for length $n$. Let $f_{(x,\omega_s)}$ be the function defined by $f_x(\langle \omega_s, y \rangle) = \langle z_1, f_{(x,\omega_s)}(y) \rangle$. Note that there exists a polynomial-time computable function $f_2$ such that $f_{(x,\omega_s)} = \lambda y.f_2(x, \omega_s, y)$. If $s - 1 > 1$, we define $(s-1)$-easy and $(s-1)$-hard strings in analogy to the above. If an $(s-1)$-hard string exists we can repeat the process and define $(s-2)$-easy and $(s-2)$-hard strings and so on. Note that the definition of $j$-easy and $j$-hard strings can only be made with respect to our hard string $x$, some fixed $s$-hard string $\omega_s$, some fixed $(s-1)$-hard string $\omega_{s-1}$, ..., some fixed $(j+1)$-hard string $\omega_{j+1}$. If we have found a sequence of strings $(\omega_s, \omega_{s-1}, \ldots, \omega_2)$ (note that if $s=1$, $(\omega_s, \omega_{s-1}, \ldots, \omega_2)$ is the empty sequence) such that every $\omega_j$ is $j$-hard with respect to $(x, \omega_s, \omega_{s-1}, \ldots, \omega_{j+1})$ then we have for all $y$, $|y| \leq n$,

$$y \in L_{\Pi_i^p} \Leftrightarrow f_{(x,\omega_s,\omega_{s-1},\ldots,\omega_2)}(y) \notin L_{\Pi_i^p}.$$

We say that a string $y$ is 1-easy for length $n$ if and only if $|y| \leq n$ and $f_{(x,\omega_s,\omega_{s-1},\ldots,\omega_2)}(y) \notin L_{\Pi_i^p}$. We define that no string is 1-hard for length $n$.

$(x)$ is called a hard sequence for length $n$ if and only if $x$ is hard for length $r'(n)$. A sequence $(x, \omega_s, \omega_{s-1}, \ldots, \omega_j)$ of strings is called a hard sequence for length $n$ if and only if $x$ is hard for length $r'(n)$ and for all $\ell$, $j \leq \ell \leq s$, $\omega_\ell$ is $\ell$-hard for length $n$ with respect to $(x, \omega_s, \omega_{s-1}, \ldots, \omega_{\ell+1})$. Note that given a hard sequence $(x, \omega_s, \omega_{s-1}, \ldots, \omega_j)$ for length $n$, the strings in $(L_{\Pi_i^p})^{\leq n}$ divide into $(j-1)$-easy and $(j-1)$-hard strings (with respect to $(x, \omega_s, \omega_{s-1}, \ldots, \omega_j)$) for length $n$.

$(x)$ is called a maximal hard sequence for length $n$ if and only if $(x)$ is a hard sequence for length $n$ and there exists no $s$-hard string for length $n$. A hard sequence $(x, \omega_s, \omega_{s-1}, \ldots, \omega_j)$ for length $n$ is called a maximal hard sequence for length $n$ if and only if there exists no $(j-1)$-hard string for length $n$ with respect to $(x, \omega_s, \omega_{s-1}, \ldots, \omega_j)$. If we in the following denote a maximal hard sequence by $(x, \omega_s, \omega_{s-1}, \ldots, \omega_j)$ we explicitly include the case that the maximal hard sequence might be $(x)$ or $(x, \omega_s)$.

**Claim 1:** *There exists a set $A \in \Sigma_i^p$ such that if $(x, \omega_s, \omega_{s-1}, \ldots, \omega_j)$ is a maximal hard sequence for length $n$, then for all $y$ and $n$ satisfying $|y| \leq n$ it holds that:*

$$y \in L_{\Pi_i^p} \Leftrightarrow \langle x, 1^n, \omega_s, \omega_{s-1}, \ldots, \omega_j, y \rangle \in A.$$

*Proof of Claim 1:* Let $(x, \omega_s, \omega_{s-1}, \ldots, \omega_j)$ be a maximal hard sequence for length $n$. If $(x, \omega_s, \omega_{s-1}, \ldots, \omega_j) = (x)$, we let $j = s+1$. Note that $j \geq 2$ and that the strings in $(L_{\Pi_i^p})^{\leq n}$ are exactly the strings of length at most $n$ that are $(j-1)$-easy with respect to $(x, \omega_s, \omega_{s-1}, \ldots, \omega_j)$. It is immediate from the definition that testing whether a string $y$ is $(j-1)$-easy for length $n$ with respect to $(x, \omega_s, \omega_{s-1}, \ldots, \omega_j)$ can be done by a $\Sigma_i^p$ algorithm running in time polynomial in $n$: If $j \geq 3$, check that $|y| \leq n$, guess $y_2$, $|y_2| \leq t^{(j-3)}(n)$, compute $f_{(x,\omega_s,\omega_{s-1},\ldots,\omega_j)}(\langle y, y_2 \rangle) = \langle z_1, z_2 \rangle$, and accept if and only if $z_1 \notin L_{\Pi_i^p}$; If $j = 2$, check $|y| \leq n$, and accept if and only if $f_{(x,\omega_s,\omega_{s-1},\ldots,\omega_2)}(y) \notin L_{\Pi_i^p}$.

**Claim 2:** *There exist a set $B \in \Sigma_i^p$ and a polynomial $\widehat{p}$ such that $(\forall n \geq 0)[\widehat{p}(n+1) > \widehat{p}(n) > 0]$ and if $(x, \omega_s, \omega_{s-1}, \ldots, \omega_j)$ is a maximal hard sequence for length $\widehat{p}(n)$, then for all $y$ and $n$ satisfying*

$|y| \leq n$ it holds that:
$$y \in \widetilde{L}_{\Sigma_{i+1}^p} \Leftrightarrow \langle x, 1^{\widehat{p}(n)}, \omega_s, \omega_{s-1}, \ldots, \omega_j, y \rangle \in B.$$

*Proof of Claim 2:* Let $A \in \Sigma_i^p$ as in Claim 1. Let $y$ be a string such that $|y| \leq n$. According to the definition of $\widetilde{L}_{\Sigma_{i+1}^p}$,
$$y \in \widetilde{L}_{\Sigma_{i+1}^p} \Leftrightarrow (\exists z \in \Sigma^{|y|})[\langle y, z \rangle \notin L_{\Sigma_i^p}].$$

Recall that $L_{\Pi_i^p} = \overline{L_{\Sigma_i^p}}$ from Notation 4.2. Define $\widehat{p}$ to be a polynomial such that $\widehat{p}(n+1) > \widehat{p}(n) > 0$ and $\widehat{p}(n) \geq t(n)$ for all $n$. Applying Claim 1 we obtain that if $(x, \omega_s, \omega_{s-1}, \ldots, \omega_j)$ is a maximal hard sequence for length $\widehat{p}(n)$, then
$$y \in \widetilde{L}_{\Sigma_{i+1}^p} \Leftrightarrow (\exists z \in \Sigma^{|y|})[\langle x, 1^{\widehat{p}(n)}, \omega_s, \omega_{s-1}, \ldots, \omega_j, \langle y, z \rangle \rangle \in A].$$

We define $B$ to be the set $B = \{\langle x, 1^{\widehat{p}(n)}, \omega_s, \omega_{s-1}, \ldots, \omega_j, y \rangle \mid (\exists z \in \Sigma^{|y|})[\langle x, 1^{\widehat{p}(n)}, \omega_s, \omega_{s-1}, \ldots, \omega_j, \langle y, z \rangle \rangle \in A]\}$. Clearly $B \in \Sigma_i^p$. This proves Claim 2.

**Claim 3:** *There exist a set $C \in \Sigma_{i+1}^p$ and a polynomial $\widehat{p}_1$ such that $(\forall n \geq 0)[\widehat{p}_1(n+1) > \widehat{p}_1(n) > 0]$ and if $(x, \omega_s, \omega_{s-1}, \ldots, \omega_\ell)$ is a maximal hard sequence for length $\widehat{p}_1(n)$, then for all $y$ and $n$ satisfying $|y| \leq n$ it holds that:*
$$y \in L_{\Sigma_{i+2}^p}^\dagger \Leftrightarrow \langle x, 1^n, \omega_s, \omega_{s-1}, \ldots, \omega_\ell, y \rangle \in C.$$

*Proof of Claim 3:* Let $B \in \Sigma_i^p$ and $\widehat{p}$ be a polynomial, both as defined in Claim 2. Let $y$ be a string such that $|y| \leq n$. According to the definition of $L_{\Sigma_{i+2}^p}^\dagger$,
$$y \in L_{\Sigma_{i+2}^p}^\dagger \Leftrightarrow (\exists z \in \Sigma^{|y|})[\langle y, z \rangle \notin \widetilde{L}_{\Sigma_{i+1}^p}].$$

Define $\widehat{p}_1$ to be a polynomial such that $\widehat{p}_1(n+1) > \widehat{p}_1(n) > 0$ and $\widehat{p}_1(n) \geq \widehat{p}(t(n))$ for all $n$. Applying Claim 2, we obtain that if $(x, \omega_s, \omega_{s-1}, \ldots, \omega_\ell)$ is a maximal hard sequence for length $\widehat{p}_1(n)$, then
$$y \in L_{\Sigma_{i+2}^p}^\dagger \Leftrightarrow (\exists z \in \Sigma^{|y|})[\langle x, 1^{\widehat{p}_1(n)}, \omega_s, \omega_{s-1}, \ldots, \omega_\ell, \langle y, z \rangle \rangle \notin B].$$

Let $C = \{\langle x, 1^n, \omega_s, \omega_{s-1}, \ldots, \omega_j, y \rangle \mid (\exists z \in \Sigma^{|y|})[\langle x, 1^{\widehat{p}_1(n)}, \omega_s, \omega_{s-1}, \ldots, \omega_j, \langle y, z \rangle \rangle \notin B]\}$. Clearly $C \in \Sigma_{i+1}^p$.

We are now ready to prove the claim of Lemma 4.3. Note that the set
$$E = \{\langle x, 1^n, \omega_s, \omega_{s-1}, \ldots, \omega_j \rangle \mid \text{for all } \ell, j \leq \ell \leq s,$$
$$\omega_\ell \text{ is } \ell\text{-hard for length } n \text{ with respect to } (x, \omega_s, \omega_{s-1}, \ldots, \omega_{\ell+1})\}$$

is in $\Pi_i^p$. Consequently, the set

$$F = \{\langle x, 1^n, k\rangle \mid (\exists\, \omega_s, \omega_{s-1}, \ldots, \omega_{s-k+2})[\langle x, 1^n, \omega_s, \omega_{s-1}, \ldots, \omega_{s-k+2}\rangle \in E]\}$$

is in $\Sigma_{i+1}^p$. Observe that if $x$ is a hard string for length $r'(\widehat{p}_1(n))$, then $\langle x, 1^n, \omega_s, \omega_{s-1}, \ldots, \omega_j\rangle \in E$ if and only if $(x, \omega_s, \omega_{s-1}, \ldots, \omega_j)$ is a hard sequence for length $\widehat{p}_1(n)$. Similarly, if $x$ is a hard string for length $r'(\widehat{p}_1(n))$, then $\langle x, 1^n, k\rangle \in F$ if and only if there exists a hard sequence (starting with $(x, \ldots)$) *of length $k$* for length $\widehat{p}_1(n)$.

It follows from those observations and the above proven claims that if $x$ is a hard string for length $r'(\widehat{p}_1(n))$, then the following algorithm will accept $\langle x, 1^n, y\rangle$ if and only if $y \in L_{\Sigma_{i+2}^p}^\dagger$. On input $\langle x, 1^n, y\rangle$ the algorithm proceeds as follows:

1. Using $F$ as an oracle, compute the largest $k$, call it $\widehat{k}$, such that $\langle x, 1^n, k\rangle \in F$.

2. Then, by making one oracle query, check the following: Do there exist strings $\omega_s, \omega_{s-1}, \ldots, \omega_{s-\widehat{k}+2}$ such that $\langle x, 1^n, \omega_s, \omega_{s-1}, \ldots, \omega_{s-\widehat{k}+2}\rangle \in E$ and $\langle x, 1^n, \omega_s, \omega_{s-1}, \ldots, \omega_{s-\widehat{k}+2}, y\rangle \in C$? Though we actually are allowed as many queries as we like (within our time bound), we note that it is not hard to see that this checking can be done by making one query to an appropriately chosen $\Sigma_{i+1}^p$ oracle.

3. Accept if and only if the final query returned the answer "yes."

Though the above algorithm queries two different $\Sigma_{i+1}^p$ oracles it is clearly a $\mathrm{P}^{\Sigma_{i+1}^p}$ algorithm, since $\Sigma_{i+1}^p$ is closed under disjoint union. Let $D$ be the set accepted by this algorithm. Define $r$ to be the polynomial such that $r(n) = r'(\widehat{p}_1(n))$ for all $n$. Note that due to the definitions of $r'$ and $\widehat{p}_1$, $r$ satisfies $r(n+1) > r(n) > 0$ for all $n$. This completes the proof of Lemma 4.3. ∎

## 5 Conclusions

We have proven a general downward translation of equality, Theorem 3.3, sufficient to yield, as a corollary:

**Corollary 5.1** *For each $m > 0$ and each $k > 1$ it holds that:*

$$\mathrm{P}_{m\text{-tt}}^{\Sigma_k^p} = \mathrm{P}_{m+1\text{-tt}}^{\Sigma_k^p} \;\Rightarrow\; \mathrm{DIFF}_m(\Sigma_k^p) = \mathrm{coDIFF}_m(\Sigma_k^p).$$

The corollary follows immediately from Theorem 3.3, Proposition 2.2, and Observation 2.3. Corollary 5.1 itself has an interesting further consequence. From this corollary, it follows that for a number of previously missing cases (namely, when $m > 1$ and $k = 2$), the hypothesis $\text{P}^{\Sigma_k^p}_{m\text{-tt}} = \text{P}^{\Sigma_k^p}_{m+1\text{-tt}}$ implies that the polynomial hierarchy collapses to about one level lower in the boolean hierarchy over $\Sigma_{k+1}^p$ than could be concluded from previous papers. This is because we can, thanks to Corollary 5.1, when given $\text{P}^{\Sigma_k^p}_{m\text{-tt}} = \text{P}^{\Sigma_k^p}_{m+1\text{-tt}}$, invoke the powerful collapses of the polynomial hierarchy that are known to follow from $\text{DIFF}_m(\Sigma_k^p) = \text{coDIFF}_m(\Sigma_k^p)$. Regarding what collapses do follow from $\text{DIFF}_m(\Sigma_k^p) = \text{coDIFF}_m(\Sigma_k^p)$, a long line of research started by Kadin and Wagner a decade ago has studied that, and the strongest currently known connection was recently obtained, independently, by Hemaspaandra, Hemaspaandra, and Hempel [12] and by Reith and Wagner [20], namely: For all $m > 0$ and all $k > 0$, if $\text{DIFF}_m(\Sigma_k^p) = \text{coDIFF}_m(\Sigma_k^p)$, then $\text{PH} = \text{DIFF}_m(\Sigma_k^p)\mathbf{\Delta}\text{DIFF}_{m-1}(\Sigma_{k+1}^p)$. Putting all the above together, one sees that, for all cases where $m > 1$ and $k > 1$, $\text{P}^{\Sigma_k^p}_{m\text{-tt}} = \text{P}^{\Sigma_k^p}_{m+1\text{-tt}}$ implies that the polynomial hierarchy collapses to $\text{DIFF}_m(\Sigma_k^p)\mathbf{\Delta}\text{DIFF}_{m-1}(\Sigma_{k+1}^p)$. And this also yields that, for all cases where $m > 1$ and $k > 1$, $\text{P}^{\Sigma_k^p[m]} = \text{P}^{\Sigma_k^p[m+1]}$ implies that the polynomial hierarchy collapses to $\text{DIFF}_{2^m-1}(\Sigma_k^p)\mathbf{\Delta}\text{DIFF}_{2^m-2}(\Sigma_{k+1}^p)$. Of course, for the case $m = 1$, we already know [13,4] that, for $k > 1$, if $\text{P}^{\Sigma_k^p[1]} = \text{P}^{\Sigma_k^p[2]}$ (equivalently, if $\text{P}^{\Sigma_k^p}_{1\text{-tt}} = \text{P}^{\Sigma_k^p}_{2\text{-tt}}$), then $\Sigma_k^p = \Pi_k^p = \text{PH}$.

**Acknowledgments** Earlier versions of parts of this paper have appeared in the Sixth Italian Conference on Theoretical Computer Science (ICTCS '98) and the Sixteenth Annual Symposium on Theoretical Aspects of Computer Science (STACS '99). We thank the referees for helpful comments and suggestions, and we thank the students of CSC486 at the University of Rochester for their proofreading help.